\documentclass[aps,prl,twocolumn,floatfix,superscriptaddress,showpacs]{revtex4}

\usepackage{graphicx}
\usepackage{dcolumn}
\usepackage{bm}
\usepackage{textcomp}

\begin{document}

\title{Specific Heat of a Fractional Quantum Hall System}


\author{F.~Schulze-Wischeler}
 \altaffiliation[current address: ]{Laboratorium f\"ur Nano- und
Quantenengineering, Gottfried Wilhelm Leibniz Universit\"at
Hannover,
  Schneiderberg 32, 30167 Hannover, Germany. }
\affiliation{ Institut f\"ur Festk\"orperphysik, Gottfried Wilhelm
Leibniz Universit\"at Hannover, Appelstra\ss{}e 2, 30167 Hannover,
Germany}

\author{U.~Zeitler}
\affiliation{
High Field Magnet Laboratory, Institute for Molecules and Materials, Radboud University Nijmegen, Toernooiveld 7, 6525 ED Nijmegen, The Netherlands
 }

\author{C.~v.~Zobeltitz}
\affiliation{
 Institut f\"ur Theoretische Physik, Gottfried Wilhelm
Leibniz Universit\"at Hannover, Appelstra\ss{}e 2, 30167 Hannover,
Germany
 }
\author{F.~Hohls}
\affiliation{ Institut f\"ur Festk\"orperphysik, Gottfried Wilhelm
Leibniz Universit\"at Hannover, Appelstra\ss{}e 2, 30167 Hannover,
Germany}

\author{D.~Reuter}
\affiliation{ Lehrstuhl f\"ur Angewandte Festk\"orperphysik,
Ruhr-Universit\"at Bochum, Universit\"atsstra\ss{}e 150, 44780
Bochum, Germany
 }
\author{A.D.~Wieck}
 \affiliation{
Lehrstuhl f\"ur Angewandte Festk\"orperphysik, Ruhr-Universit\"at
Bochum, Universit\"atsstra\ss{}e 150, 44780 Bochum, Germany
 }
 \author{H.~Frahm}
\affiliation{
 Institut f\"ur Theoretische Physik, Gottfried Wilhelm
Leibniz Universit\"at Hannover, Appelstra\ss{}e 2, 30167 Hannover,
Germany
 }
 \author{R.J.~Haug}
\affiliation{ Institut f\"ur Festk\"orperphysik, Gottfried Wilhelm
Leibniz Universit\"at Hannover, Appelstra\ss{}e 2, 30167 Hannover,
Germany}

\date{\today}

\begin{abstract}
Using a time-resolved phonon absorption technique, we have
measured the specific heat of a two-dimensional electron system in
the fractional quantum Hall effect regime. For filling factors
$\nu = 5/3, 4/3, 2/3, 3/5, 4/7, 2/5$ and 1/3 the specific heat
displays a strong exponential temperature dependence in agreement
with excitations across a quasi-particle gap. At filling factor
$\nu = 1/2$ we were able to measure the specific heat of a
composite fermion system for the first time. The observed linear
temperature dependence on temperature down to $T = 0.14$~K agrees
well with early predictions for a Fermi liquid of composite
fermions.

\end{abstract}

\maketitle


In the fractional quantum Hall (FQH) effect the Coulomb
interaction induces the formation of new quasi-particle states
characterized by a fractional charge and a finite excitation
gap~\cite{Stormer,Laughlin,book1,book2}. A nice description of the
FQH effect is given by the composite fermion (CF)
theory~\cite{Jain_89} where the FQH effect is regarded as an
integer quantum Hall effect of new quasi particles. In particular,
at a filling factor $\nu = 1/2$ these composite fermions can be
described as a quasi-free Fermi sea in an effective zero-magnetic
field~\cite{Halperin_93}.

One of the experimental challenges lies in accessing the ground
state properties of a two-dimensional electron system (2DES) by
means of thermodynamic quantities such as
magnetization~\cite{Eisenstein, Wiegers, Meinel}, thermal
conductivity~\cite{Eisenstein_87, Bayot_92},
thermopower~\cite{Zeitler_92, Tieke_96} or specific heat
\cite{Gornik, Bayot_03}. It is, however, far from being
straightforward to measure the specific heat $C$ of a 2DES
directly. In general $C$ is strongly dominated by the contribution
of the surrounding substrate, and, in order to obtain a reasonable
signal from the 2DES, multi-layer structures have to be
used~\cite{Gornik, Bayot_03}.

An alternative method lies in the use of time resolved
phonon-spectroscopy experiments
\cite{Mellor_95_First_Phonons,Kent_96}. Here, a defined amount of
energy is dissipated inside the 2DES in a short period of time
(typically a few nanoseconds). Measuring the temperature of the
2DES long before it reaches equilibrium with the substrate then
allows to determine its specific heat directly~\cite{Zeitler_99}.

In this letter we will show how we can use phonon-absorption
experiments to deduce the temperature dependence of $C$ of a 2DES
in the FQH regime. Results at fractional filling factors suggest
the excitation across a quasi-particle gap in quantitative
agreement with theoretical predictions~\cite{Chakraborty_97}. At
filling factor $\nu = 1/2$ we essentially find a linear
temperature dependence predicted for a Fermi liquid of composite
fermions~\cite{Halperin_93}. Slight deviations from linearity will
be tracked down to spin-splitting in the CF system.

The sample used in this experiment is a 2DES embedded in a
modulation-doped AlGaAs/GaAs-heterostructure. Using its persistent
photoconductivity the electron concentration $n_e$ inside the 2DES
can be tuned be means of an infrared photo diode. The sample is
mounted on the cold finger of a dilution refrigerator and inserted
into a 13 T superconducting magnet. In order to heat up the 2DES
on short time scales of a few nanoseconds, it is bombarded with
ballistic phonon pulses emitted from a heater on the backside of
the 2~mm thick GaAs substrate; a more detailed description of this
phonon absortion technique can be found elsewhere~\cite{Kent_96,
Zeitler_99, schuwi_04}. The temperature change inside the 2DES is
measured by converting its time-dependent resistance $R$ to a
temperature $T$ using the equilibrium calibration for $R(T)$. In
order to maximize the signal to noise ratio, the 2DES is patterned
into a long meander, see Ref.~\onlinecite{schuwi_04} for more
details.

\begin{figure}[t]  
  \begin{center}
  \resizebox{1\linewidth}{!}{\includegraphics[angle=270]{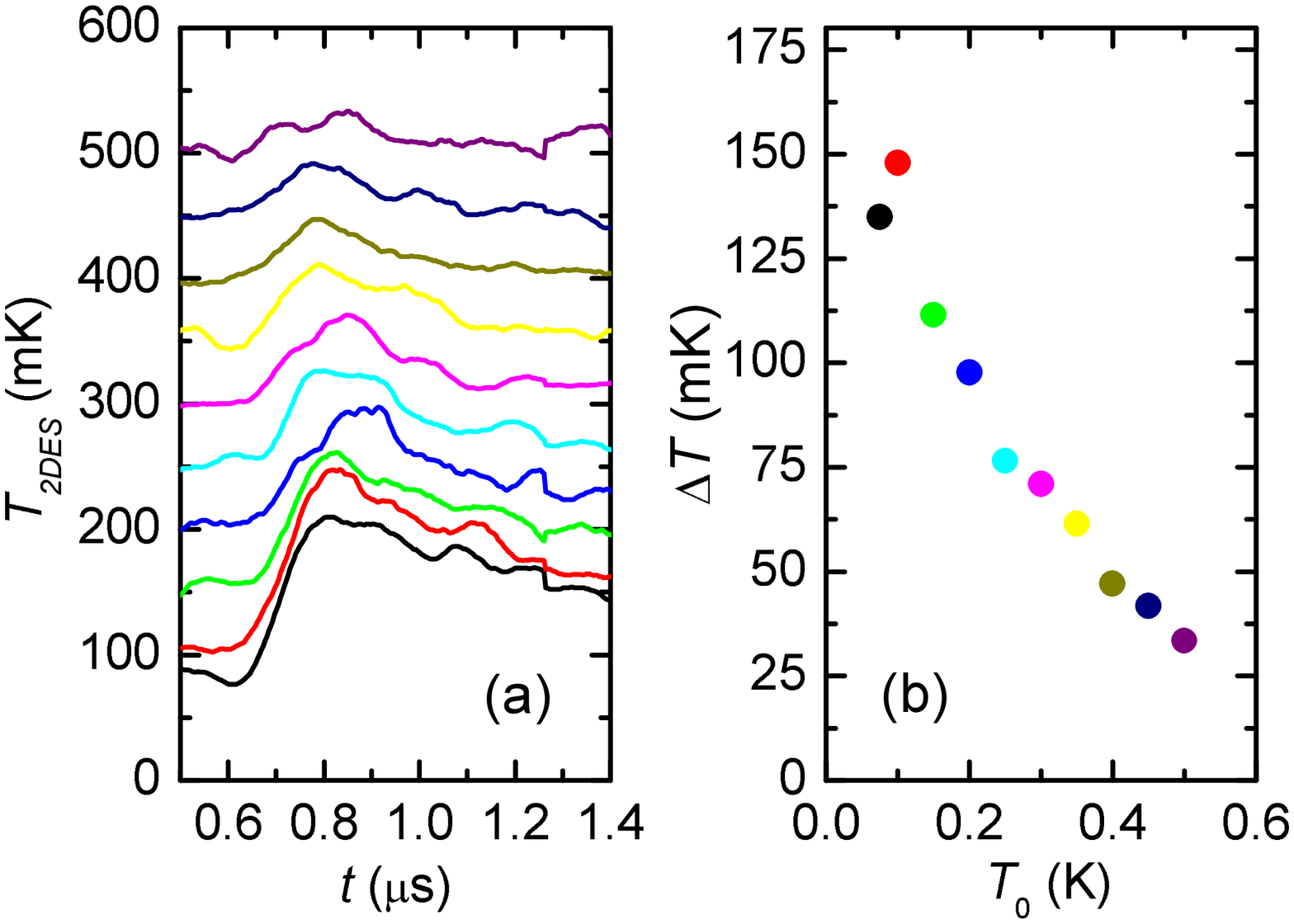}}
  \end{center}
\caption{(Color online). Phonon signal data at $\nu = 1/2$ ($n_e =
1.03\times 10^{15}$~m$^{-2}$ ). \newline (a): Phonon signal in mK
versus time for same phonon temperature $T_P = 1.9$~K and pulse
length $\tau = 10$~ns. From bottom to top and the equilibrium
temperature of the 2DES is increased from 75 mK to 500 mK.
\newline
(b): Total temperature change $\Delta T$ of the 2DES due to
ballistic-phonon absorption as a function of $T_0$. }
  \label{Figure_PS_dT.eps}
\end{figure}

Fig.~\ref{Figure_PS_dT.eps}(a) shows a typical phonon signal for a
2DES at filling factor $\nu = 1/2$ at different equilibrium
temperatures $T_0$ between 75~mK and 500~mK. At $t=0$ a 10~ns
phonon pulse is emitted. This pulse is characterized by a
non-equilibrium phonon temperature $T_P = 1.9$~K. After 0.6 $\mu
s$ the phonon pulse reaches the 2DES which leads to an increase of
its temperature to a maximum peak height $T_1$. The magnitude of
the phonon signal, $\Delta T = T_1 - T_0$,  decreases with
increased $T_0$.  This fact is emphasized in
Fig.~\ref{Figure_PS_dT.eps}(b) where we have plotted $\Delta T$ as
a function of $T_0$. As we will quantify further on the monotonic
decrease of $\Delta T$ is directly related to a monotonic increase
of the 2DES specific heat at $\nu = 1/2$.

In order to extract the 2DES specific heat from this phonon
absorption experiments we introduce a simple model: The
temperature change $dT$ of a 2DES due to phonon absorption can be
written as:
\begin{equation}\label{model1}
  C(T)dT = r(T,T_P)P_P dt
\end{equation}
where $C(T)$ is the 2DES's specific heat, $r(T,T_P)$ is an
absorption coefficient depending on the non-equilibrium phonon
temperature $T_P$ and 2DES temperature $T$. $P_P$ is total power
inside one phonon pulse. Since phonon emission cooling the 2DES
back to its equilibrium temperature takes place at much longer
timescales (typically $\mu s)$ it can be neglected on the 10-ns
timescale for phonon absorption. Integrating this general equation
over the phonon-pulse length $\tau$ leads to
\begin{equation}\label{model2}
  \int_{T_0}^{T1}C(T)dT =  r_0 P_P \tau
\end{equation}

In the experiment the phonon temperature $T_P$ is kept constant,
the absorption coefficient $r(T,T_P)$ only depends on the
temperature $T$. For low enough temperatures $T << T_P$ the 2DES
will be essentially in its (temperature independent) ground state
and $r$ can be regarded as constant. This is indeed the case for
our experiments which justifies the use of a roughly constant
absorption rate $r_o$ for all temperatures used.


\begin{figure}[t]  
  \begin{center}
  \resizebox{1\linewidth}{!}{\includegraphics{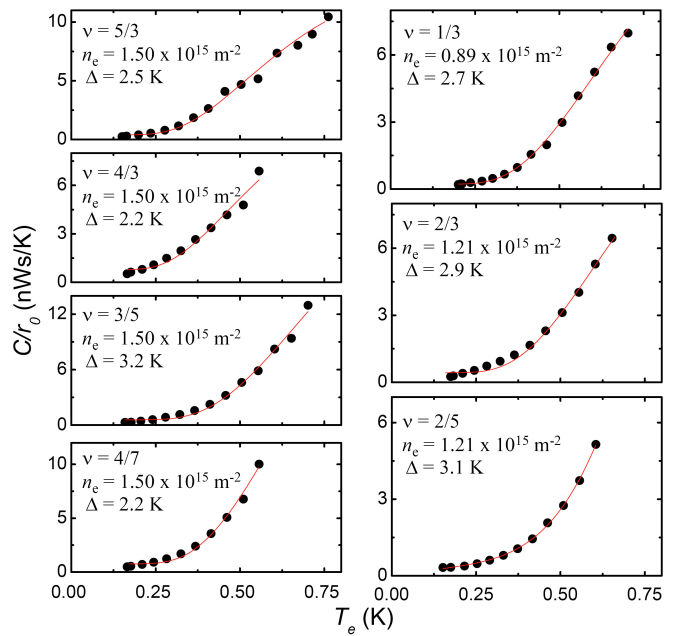}}
  \end{center}
\caption{(Color online). Relative specific heat $C(T)/r_0$ for the
filling factors $\nu = 5/3, 4/3, 2/3, 3/5, 4/7, 2/5$, and $1/3$
measured with a phonon absorption experiment. The unknown
absorption coefficient $r_0$, $0 < r_0 < 1$, is fixed for every
single graph. The lines are fits with the theory (see text).}
  \label{Figure_C_other.eps}
\end{figure} 

Using Eq.~(\ref{model2}) we can then determine the (relative)
specific heat of the 2DES, $C(T)/r_0$, from a set of experiments
as shown in Fig.~1. Although $r_0$ is a-priory unknown this
procedure already allows a direct access to the temperature
dependence of the specific heat of a 2DES.

Results for the specific heat of FQH filling factors $\nu = 5/3,
4/3, 2/3, 3/5, 4/7, 2/5$, and $1/3$ are shown in
Fig.~\ref{Figure_C_other.eps}, the corresponding electron
concentration $n_e$ are marked in the figure. Please note that the
measurements were taken at different $n_e$ because filling factors
5/3, 4/3, 3/5, and 4/7 showed up only at highest electron
concentration (due to higher electron mobility of the illuminated
sample) whereas $\nu = 1/3$ was only accessible for lower electron
concentrations in our magnet, 2/3 und 2/5 shows data at a medium
electron concentration. All the curves are taken for different
electron concentrations show the same characteristic form: for low
temperatures $T$ the relative specific heat $C(T)/r_{0}$ is flat
and finite, and increases exponentially for higher temperatures.

The theoretically predicted specific heat of such a system is $C
\propto (1/T)^2 exp(-\Delta/T)$ \cite{Chakraborty_97}. Here
$\Delta$ is the energy gap from the occupied ground state to the
following unoccupied Landau level of composite fermions. The solid
lines shown in Fig.~\ref{Figure_C_other.eps} show a comparison of
this theory with our experimental data. We have added a small
empirical constant caused by a finite (thermodynamic) density of
states inside the excitation gap \cite{Zeitler_99}. The energy
gaps $\Delta$ deduced from this procedure [see
Fig.~\ref{Figure_C_other.eps}] are comparable to gaps determined
from temperature dependent transport experiments~\cite{schuwi_04,
remark}, but show a large uncertainty due to fits far away from
the temperatures corresponding to the gab energies, especially at
$\nu = 1/3$.

\begin{figure*}[t]  
  \begin{center}
  \resizebox{0.55\linewidth}{!}{\includegraphics{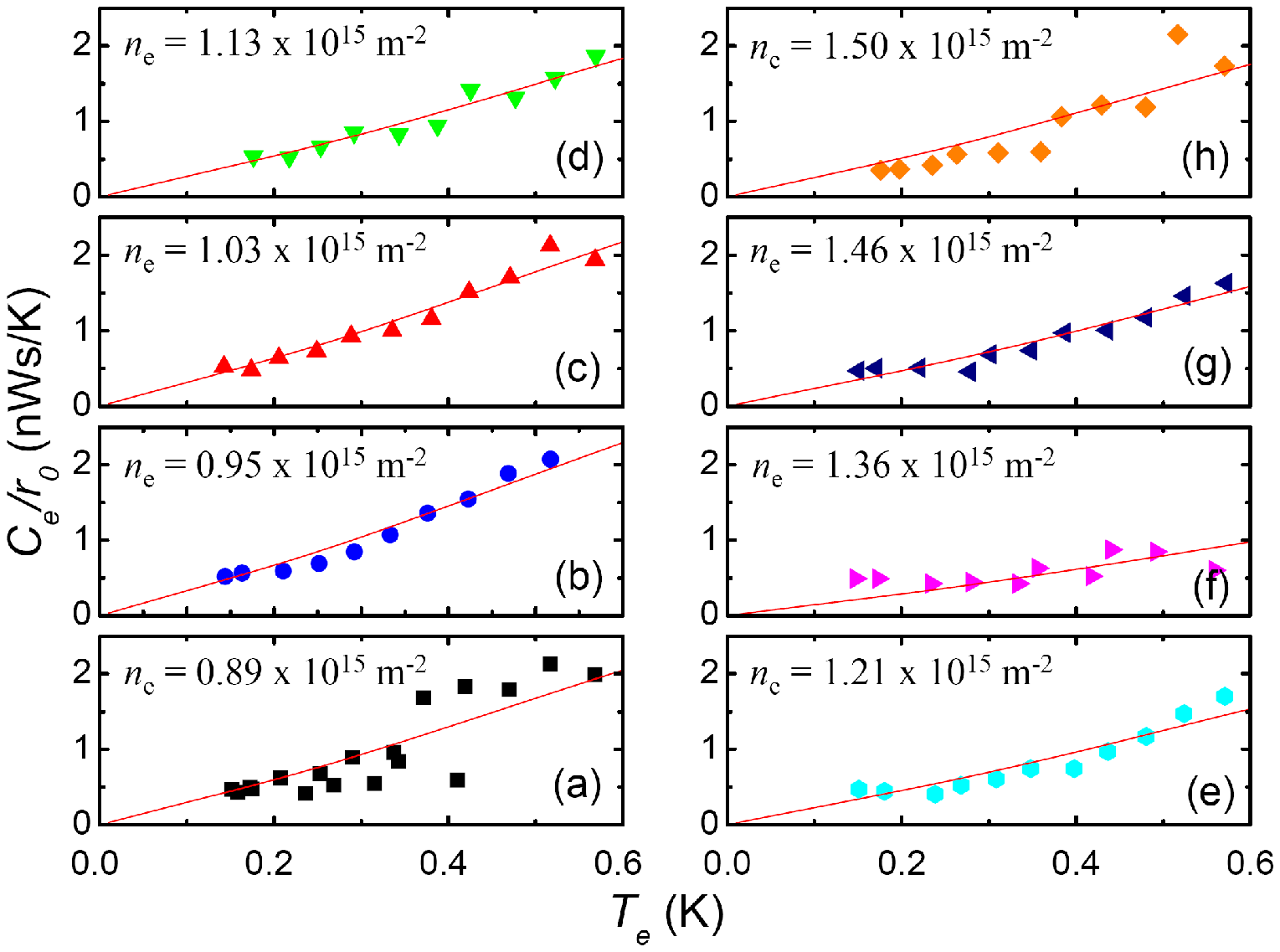}}
  \end{center}
\caption{(Color online). Relative specific heat $C(T)/r_0$ at
filling factor $\nu = 1/2 $ for eight different electron densities
from $n_e = 0.89\times 10^{15}$~m$^{-2}$(a) up to $n_e =
1.50\times 10^{15}$~m$^{-2}$ (h) measured in a phonon absorption
experiment. The lines are fits with the theoretical expected
linear behavior through the origin \cite{Halperin_93} plus a term
$(\Delta_{spin}^2 / T) \, e^{-  \Delta_{spin} /k_B T}$ taking the
influence of a second constant density of states into account. The
error of the data points is given by their distribution.}
  \label{Figure_C_1d2.eps}
\end{figure*} 

The temperature dependence of the specific heat at $\nu=1/2$ is
distinctively different from the fractional filling factors shown
above.  Fig.~\ref{Figure_C_1d2.eps} shows our results for relative
specific heat $C(T)/r_0$ at $\nu = 1/2$ for eight different
electron densities ranging from $n_e = 0.89 \times
10^{15}$~m$^{-2}$ to $n_e = 1.50 \times 10^{15}$~m$^{-2}$. Note
that all measurements are done with the same standard phonon pulse
($\tau = 10$~ns, $T_P = 1.9$~K). Since the phonon signal at $\nu =
1/2$ is extremely weak we had to average over up to 1.5 million
individual traces in order to obtain reliable data. The reason for
the weakness of the signal (and therefore the large amount of
scatter in the data) is that the resistance nearly not changes as
a function of temperature at $\nu = 1/2$. For higher temperatures
the resistance change goes nearly to zero and for this physical
reason the specific heat at $T > 0.6$ K can not be measured with
phonon absorption experiments here at all. The leading temperature
dependence of the specific heat for small $T$ is found to be
linear.  This agrees with the CF picture of the FQHE in mean-field
approximation (i.e.\ neglecting fluctuations of the fictitious
gauge field introduced in the transformation to CFs) as a
consequence of quasi-particle and quasi-hole close excitations
close to the Fermi surface~\cite{Halperin_93}. Our experimental
observation therefore form another strong confirmation of the CF
picture of the FQHE.

The dependence of the relative specific heat $C(T)/r_0$ on
electron concentration $n_e$ is governed by two effects. On the
one hand, the composite-fermion mass $m_{CF}$ increases
proportional to $\sqrt{n_e}$ \cite{Halperin_93,Jain_97}. On the
other hand, the coupling of TA-phonons to a 2DES is piezoelectric
and therefore proportional to the electron density $n_e$
\cite{Kent_96,Benedict_99}, i. e. $r_0 \propto n_e$. These two
competing processes lead to a concentration  dependence of the
relative specific heat $C(T)/r_0 \propto 1 / \sqrt{n_e}$
consistent with our experimental data.

Beyond the leading linear temperature dependence the analysis of
our data gives a small positive contribution to $C$.  This is
different from the CF picture for a single half-filled Landau
level where gauge fluctuations leading to logarithmic corrections
to the entropy are expected to reduce the specific heat
\cite{Halperin_93}.

\begin{figure}[b]  
  \begin{center}
  \resizebox{0.8\linewidth}{!}{\includegraphics{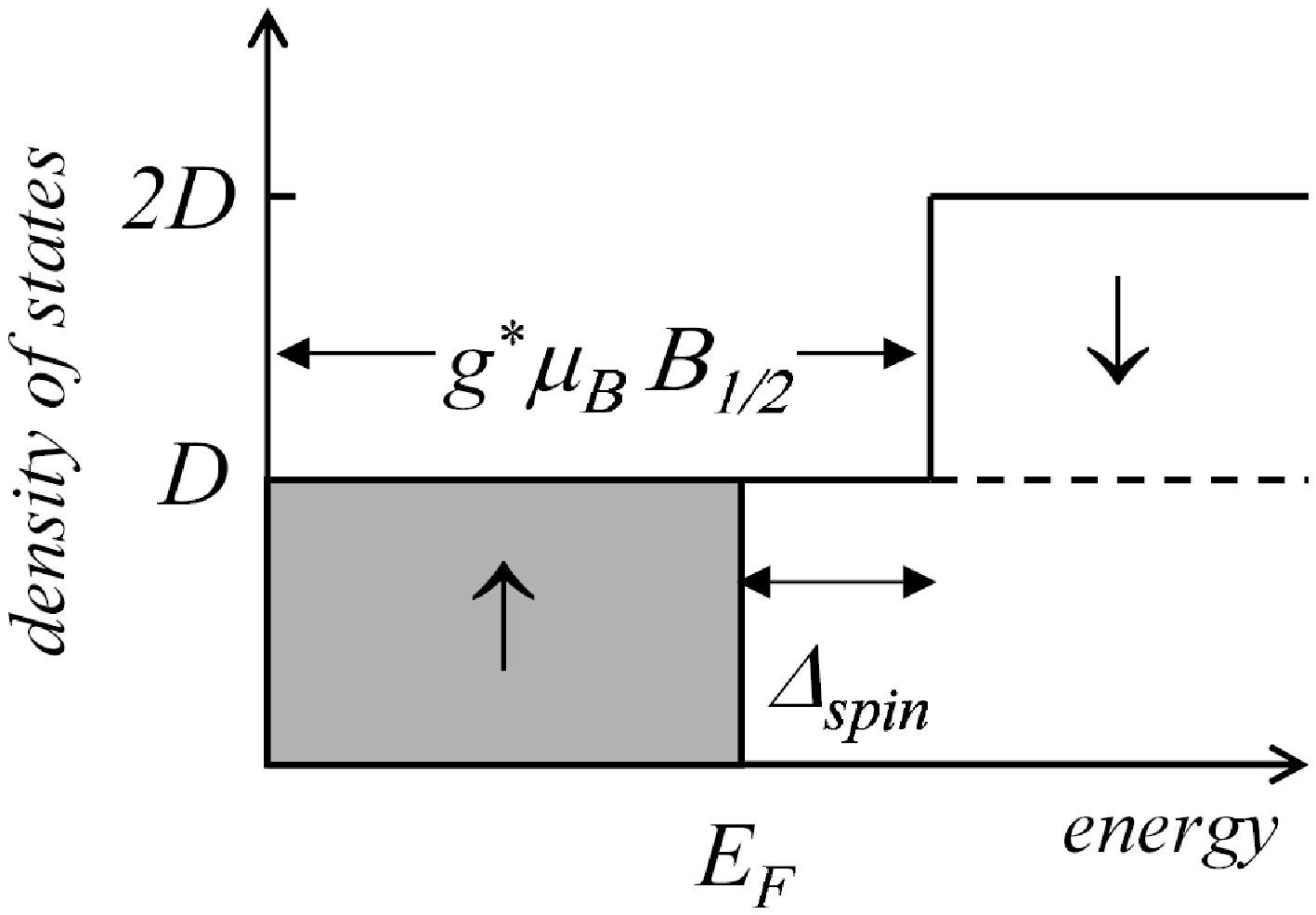}}
  \end{center}
\caption{Model for the density of states at $\nu = 1/2$ including
spin. At $T = 0$ the system is spin polarized and filled with
spin-up CF up to the Fermi energy $E_F$. The second (spin-down)
band is shifted by the Zeeman splitting $g^* \mu_B B_{1/2}$, the
lowest lying separated from $E_F$ by an energy gap
$\Delta_{spin}$.}
  \label{model.eps}
\end{figure} 

To explain this behavior we propose a description taking into
account the spin of the CF~\cite{Kramer_2002,Mariani_2002}: at the
magnetic field $B_{1/2}$ corresponding to filling factor $\nu=1/2$
the ground state of the system is completely spin polarized. The
second Landau level for the electrons with opposite spin is
separated from the Fermi energy $E_F$ by a finite gap
$\Delta_{spin} = g^* \mu_B B_{1/2} - E_F$ (see
Fig.~\ref{model.eps}). Here $g^*$ is the effective Land\'e factor
of the CFs.

Within this picture, the total energy $E_{tot}$ of the system at a
finite temperature $T$ is
\begin{eqnarray}
\label{CvZ4}
 E_{tot} =   \int_{0}^{\infty}
 \frac{D\;E}{1 + e^{\beta (E - \mu(T))}} dE\\ \nonumber
 +  \int_{E_F+\Delta_{spin}}^{\infty}
\frac{D\;E}{1 + e^{\beta (E - \mu(T)) }} dE
\end{eqnarray}
The second term represents the contribution of the second spin state to the total energy.

with the constant density of states $D$ of the CF system.
Computing these integrals the leading terms of the specific heat
are found to be
\begin{equation}\label{CvZ6}
  C = -\frac{\partial E_{tot}}{\partial T}
  \approx \frac{\pi^2}{3} D k_B T + D \frac{\Delta_{spin}^2}{k_B T} e^{-\Delta_{spin}/k_B T}
\end{equation}

Please note that the first linear term is the known result from
Halperin, Lee and Read $C=\frac{\pi}{6}m_{CF} \, k_{B}^2 T$
\cite{Halperin_93} by inserting the DOS $D = N/E_F$ and the Fermi
energy of free fermions (CFs in our case) $E_F = \frac{\hbar^2}{2
m_{CF}} (4 \pi N)$.

We have fitted our data in Fig.~\ref{Figure_C_1d2.eps} with
Eq.~\ref{CvZ6} and find indeed a good agrement. The energy gap
$\Delta_{spin}$ determined by this procedure is $\Delta_{spin}
(\nu =1/2) = 1.7 \pm 0.3$~K, independent on the electron
concentration. Since the Fermi energy $E_F$ of the composite
Fermions decreases with increasing electron concentration $E_F
\propto 1/\sqrt{n_e}$, this means implicitly that the spin
splitting of CFs at $\nu = 1/2$, $g^* \mu_B B_{1/2} =
\Delta_{spin} + E_F$ decreases when the electron concentration is
increased.

In conclusion, we have measured the temperature dependence of the
relative specific heat of a two-dimensional electron system in the
fractional quantum Hall system at various filling factors. The
results depend strongly on the filling factor and can be
classified into two groups: Measurements at filling factors $\nu =
5/3, 4/3, 2/3, 3/5, 4/7, 2/5$ and 1/3, where the sample is in a
state with an energy gap $\Delta$, show a clear exponential
behavior. In contrast, the measured specific heat at filling
factor $\nu = 1/2$, corresponding to a Fermi sea of composite
fermions, shows a linear temperature dependence; small deviations
from this linearity could be explained with the influence of a
second spin state at $\nu = 1/2$.

We acknowledge financial support by DFG priority program ``quantum
Hall systems'' and discussions with W. Apel (PTB Braunschweig), W.
Weller (ITP Leipzig), and K. v. Klitzing (MPI-Fkf Stuttgart).



\begin{thebibliography}{99}
\bibitem{remark}
The energy gaps as mesured from the specific heat are
thermodynamic gaps susceptible to the disorder in the system. A
method using phonon absorption experiments always yields larger
absolute gaps.
\end{thebibliography}
\end{document}